# Flux and energy asymmetry in a low pressure capacitively coupled plasma discharge excited by sawtooth-like waveform – a harmonic study

**Sarveshwar Sharma[1,2], Nishant Sirse[3] and Miles M Turner[4]**

[1]Institute for Plasma Research, Gandhinagar-382428, India
[2]Homi Bhabha National Institute, Anushaktinagar, Mumbai-400094, India
[3]Institute of Science and Research and Centre for Scientific and Applied Research, IPS Academy, Indore-452012, India
[4]School of Physical Sciences and National Center for Plasma Science and Technology, Dublin City University, Dublin 9, Ireland

Email: nishantsirse@ipsacademy.org

**Abstract**

Control over plasma asymmetry in a low-pressure capacitively coupled plasma (*CCP*) discharges is vital for many plasma processing applications. In this article, using the particle-in-cell simulation technique, we investigated the asymmetry generation by a temporally asymmetric waveform (sawtooth-like) in collisionless *CCP* discharge. A study by varying the number of harmonics (*N*) contained in the sawtooth waveform is performed. The simulation results predict a non-linear increase in the plasma density and ion flux with *N* i.e., it first decreases, reaching a minimum value for a critical value of *N*, and then increases almost linearly with a further rise in *N*. The ionization asymmetry increases with *N,* and higher harmonics on the instantaneous sheath position are observed for higher values of *N*. These higher harmonics generate multiple ionization beams that are generated near the expanding sheath edge and are responsible for an enhanced plasma density for higher values of *N*. The ion energy distribution function (*IEDF*) depicts a bi-modal shape for different values of *N*.  A strong *DC* self-bias is observed on the powered electrode, and its value with respect to the plasma potential decreases with an increase in *N* due to which corresponding ion energy on the powered electrode decreases. The simulation results conclude that by changing the number of harmonics of a sawtooth-like in collisionless *CCP* discharges, the ion flux asymmetry is not generated, whereas sheath symmetry could be significantly affected and therefore a systematic variation in the ion energy asymmetry is observed. Due to an increase in the higher harmonic contents in the sawtooth waveform with *N*, a transition from broad bi-modal to narrow-shaped *IEDFs* is found*.*

## I. Introduction

Low-pressure capacitively coupled plasma (*CCP*) discharges operating in the radiofrequency (*RF*) regime are extensively used in plasma processing such as thin film depositions, plasma etching, and plasma cleaning applications [1-2]. For optimizing the plasma processes, control of vital parameters such as ion energy and ion flux is highly desired. In a single-frequency operated *CCP* discharge this is often challenging as the power deposited into both electrons and ions varies with the *RF* waveform current/ voltage amplitude [3-17]. *CCP* operated using dual-frequency (*DF*) waveforms provides a level of independent control over ion energy and flux [18-29]. In such discharges, low-frequency voltage amplitude is used to control the ion energy through the sheath and high-frequency voltage amplitude control the ion flux. However, the coupling between the multiple frequencies was observed which limits the applicability of *DF-CCPs* [29-33].

Another novel concept based on the asymmetry effect was proposed to control these vital parameters. Initially, this was performed by changing the phase angle between the *DF* waveforms [42-45]. Using this technique, a *DC* self-bias was produced even in a geometrically symmetric *CCP* discharge. In this work, for the first-time plasma series resonance (*PSR*) was observed in a geometrically symmetric *CCP* discharge [44]. It was experimentally verified that by varying the phase angle between 2 frequencies of the *DF* waveform one can control the average ion energy (almost linearly) by keeping the ion flux constant [45]. Using a similar approach, the simulation results of Donko *et al.* predict a control over the shape of the ion energy distribution function (*IEDF*) on the electrodes [44]. It was reported that the higher moments in the *IEDFs* were controlled by varying the phase angle [46]. A study of the variable driving frequency showed that the control over the mean ion energy is strongly reduced at a lower driving frequency [47]. They attributed this effect to the change in the electron heating mode and enhanced charged dynamics. Changing the electrode material so as to change the secondary electron emission is another way to generate an asymmetry in a geometrically symmetric *CCP* discharge [48]. Manipulating driving frequency and discharge voltage is an alternate approach for independent control of ion flux and ion energy [14].

In the past decade, the use of non-sinusoidal voltage waveforms has emerged as a promising way to control the asymmetry and ion flux/ion energies in *CCP* discharges [49-59]. Such waveforms are known as "Tailored Waveform" that are produced by incorporating higher harmonics in the sinusoidal waveform. The asymmetry produced by tailored waveforms is broadly classified into two categories: 1) amplitude asymmetry [60] and 2) temporal asymmetry [52-53, 61]. In amplitude asymmetry, waveforms with varying minimum and maximum voltage/current excursions are used. In temporal asymmetry varying slopes are used with similar negative and positive waveform amplitudes. In temporally asymmetric waveforms, a slope asymmetry is introduced such as saw-tooth like waveforms to generate an asymmetric plasma response. Using *PIC* simulations, Bruneau *et al* showed that the slope asymmetric waveform induces varying sheath response near the powered and grounded electrodes [52]. One sheath triggers higher heating, a higher ionization rate, and therefore a large ion flux asymmetry is induced. Depending on the different rising and falling slopes, one electrode receives more ion flux with lower ion energy. Further study showed that the ion flux asymmetry disappears at lower pressure and higher driving frequency [53]. A systematic comparison between electropositive (*Ar*) and electronegative (*CF₄*) plasma depicts an opposite response *i.e.,* the ionization maxima switches from faster sheath expansion electrode for electropositive to faster sheath contraction for electronegative plasma [62]. High-frequency sheath modulation and higher harmonics were observed in low-pressure *CCP* discharges excited by a current driven saw-tooth like waveform [57]. The study of *IEDF* by varying the current density amplitude of saw-tooth like waveform and driving frequency presented narrow *IEDFs* at lower current density amplitudes and higher driving frequencies [59].

Another different approach to control flux and energy in *CCP* discharge is the use of an external magnetic field. Recently, a new operational regime is identified in low-pressure very high frequency (VHF) *CCP* discharge in the presence of a weak uniform external magnetic field. It is shown that high plasma density and thereby high ion flux can be achieved even at a very weak external magnetic field (~ 10 G) where ion energy can also be optimized simultaneously [34-38]. It is also reported in the literature that using magnetic field asymmetry in single-frequency CCP can independently control the ion flux and ion energy [39-40]. Later Sharma *et al.* demonstrated that enhancement in ion flux and effective control of ion energy can be achieved by imposing a uniform transverse magnetic field in low-pressure single-frequency *CCP* [41].

The present study is focused on the temporal asymmetry in a symmetric *CCP* discharge excited by a sawtooth-like current waveform. We extended our previous studies to investigate the effect of the number of harmonics (*N*) on plasma density, ionization, electron heating, ion flux, and ion energy asymmetry in a low-pressure geometrically symmetric *CCP* discharge [57, 59]. One of the previous simulation studies performed in collisional *CCP* discharge by varying a number of harmonics up to 5 [52] showed ionization, density and flux asymmetry increases with *N*. We followed the earlier studies to systematically investigate the effect of *N* on plasma asymmetry in collisionless *CCP* discharge. The low-pressure operation of *CCP* discharges is typically the condition involved in many plasma processing applications. Along with the plasma asymmetry, we systematically investigated the *IEDFs* for a number of harmonics (up to 12) onto the powered and grounded electrodes, which is never reported. A variation in the *DC* self-bias induced on the powered electrode and conditions for obtaining narrow *IEDFs* is discussed.

The present manuscript is organized as follows. Section II describes the details of simulation scheme and parameters. In section III, simulation results are discussed. The summary and conclusions of the work is presented in section IV.

## II. Simulation scheme and parameters

In this work, we adapted the particle-in-cell (*PIC*) simulation method along with the Monte Carlo Collision Scheme [63-64]. The simulation is performed in one spatial dimension and 3 velocity dimensions (1D3V) using a well-tested and benchmarked code [72] which has been widely used in literature for capacitive discharges [9-10,13-15,65-71]. This code can handle both current and voltage-driven modes and we have used the earlier one for all sets of simulations. A saw-tooth like current waveform is chosen for the plasma excitation with a mathematical expression [52] given by the equation

$$J_{RF}(t) = \pm j_0 \sum_{k=1}^{N} \frac{1}{k} \sin(k\omega_{RF} t) \quad \text{--------------------} \quad (1)$$

In equation (1), $j_0$ is the current density amplitude, $\omega_{RF}$ is the fundamental angular frequency. The positive and negative sign in equation (1) resembles to "saw-tooth down" and "saw-tooth up" waveform respectively. $N$ is the number of harmonics contained in the saw-tooth waveform, which is varied from 2-12 in the present simulation. Figure 1 shows the applied current density profiles for a few different values of $N$ (1 (sinusoidal), 2, 6, and 12). As displayed in figure 1, the current profile is temporally symmetric in the case of sinusoidal waveform i.e., for $N = 1$. With an increase in the number of harmonics, the rise time (from 0 to 50 A/m$^2$) is decreasing from 12.35 ns to 2.79 ns for $N = 2$ to $N = 12$ respectively. The current waveforms are applied at the powered electrode (i.e. at $L = 0$ cm) and the grounded electrode is at $L = 6$ cm. An external capacitor is not considered in the simulations. The *DC* self-bias generated at the powered electrode is due to the temporal asymmetry, which appeared in a self-consistent manner from the simulation based on the charge imbalance at the powered and grounded electrodes.

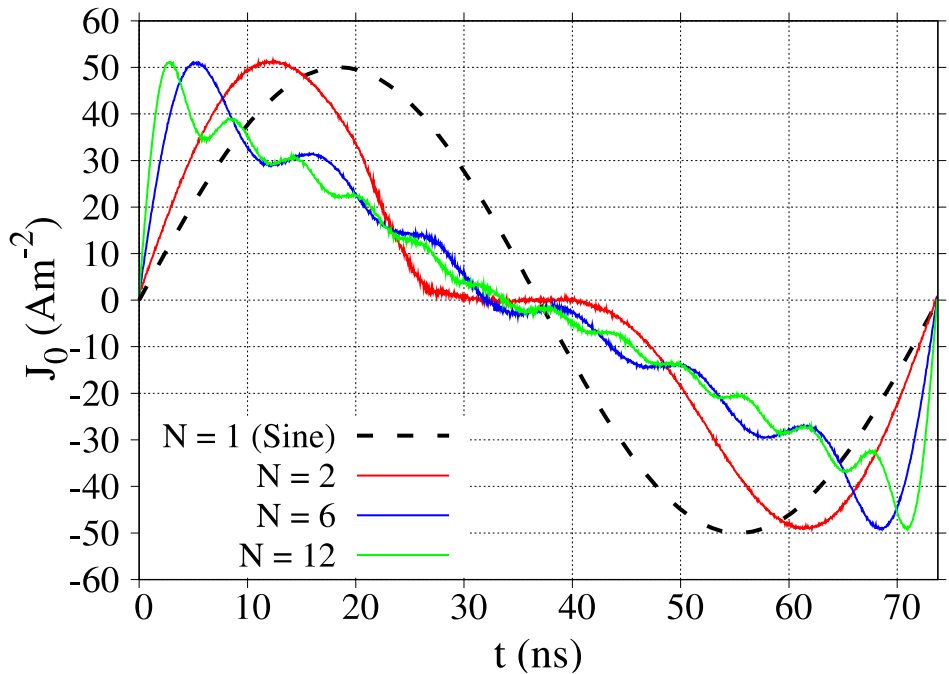

Fig.1 Applied current density profile on the powered electrode for few different values of N *i.e.*, for $N = 1, 2, 6$ and 12.

The simulation is performed for electropositive argon plasma where the gas is uniformly distributed in the discharge region at a temperature of 300 K. The gas pressure is kept constant at 5 mTorr. All the major sets of reactions including ion-neutral (elastic, inelastic and charge exchange) and electron-neutral (elastic, inelastic, and ionization) are considered here. The present simulation also considers the formation of two metastable states, Ar$^*$ and Ar$^{**}$. These two states are lumped excited states of Argon at 11.6 eV (Ar$^*$ - 3p5 4s) and 13.1 eV (Ar$^{**}$ - 3p5 4p). Other processes for the population and de-population of the excited states such as de-excitation, super-elastic collisions, metastable pooling, and multi-step ionizations are also considered in the present simulation. The collision cross-sections for all sets of reactions are taken from well-tested sources [15,67,73]. The electrodes considered here are perfectly absorbing, and electron or ion-induced secondary electron emissions (SEE) are not included. The above condition is valid for the present simulation since the operating pressure is low and the corresponding discharge voltage is also lower and therefore SEE will not play a major role in plasma dynamics [74].

To achieve the steady state solution, the simulations were performed for more than 5000 *RF* cycles. The stability and accuracy criterion of the *PIC* simulation is achieved by selecting an appropriate grid size ($\Delta x$) and time step ($\Delta t$) in order to resolve the Debye length ($\lambda_D$) and electron plasma frequency ($\omega_{pe}$) respectively [63]. The discharge gap is divided into 512 cells with 100 particles per cell. The time steps are 6400 per *RF* period. The time-averaged simulations results were averaged over 100's of *RF* period after reaching steady-state solutions.

### III.    Results and Discussions

Figure 2 (a) shows the spatial profile of time-averaged electron and ion density in the discharge for 3 different values (2, 6, and 12) of *N*. The sinusoidal density profile is also shown for comparison purposes. In the present simulations, the current density amplitude is fixed at 50 A/m$^2$ and the discharge gap is 6 cm.  As one can see, the density profile is nearly symmetric for the sinusoidal case with a maximum of ~ $8.0 \times 10^{15}$ m$^{-3}$ at the centre of the discharge. Both electron and ion density fall toward the electrodes. As *N* increases, the density profile becomes asymmetric. One can observe that the bulk plasma length is increasing with *N*, and the sheath near the powered electrode is decreasing at a much faster rate in comparison to the sheath near to the grounded electrode. Figure 2 (b) shows the systematic variation of the ratio of sheath width at the powered electrode to the grounded electrode for different values of *N* from 2-12. The plotted sheath widths are calculated as where the electron sheath edge is at the maximum from the electrode and the quasi-neutrality condition breaks down. As shown in figure 2 (b), the ratio of powered to grounded electrode sheath decreases almost linearly, from 0.75 to 0.55 with an increase in the value of *N*. The ratio is less than unity since the grounded sheaths are bigger in comparison to the powered electrode sheaths. For a given range of *N* from 2-12, the simulation results predict a total of ~30 % drop in the sheath ratio. These results suggest that the asymmetry (either in flux or energy, which is discussed later in the manuscript) in the discharge is increasing with an increase in the number of harmonics.

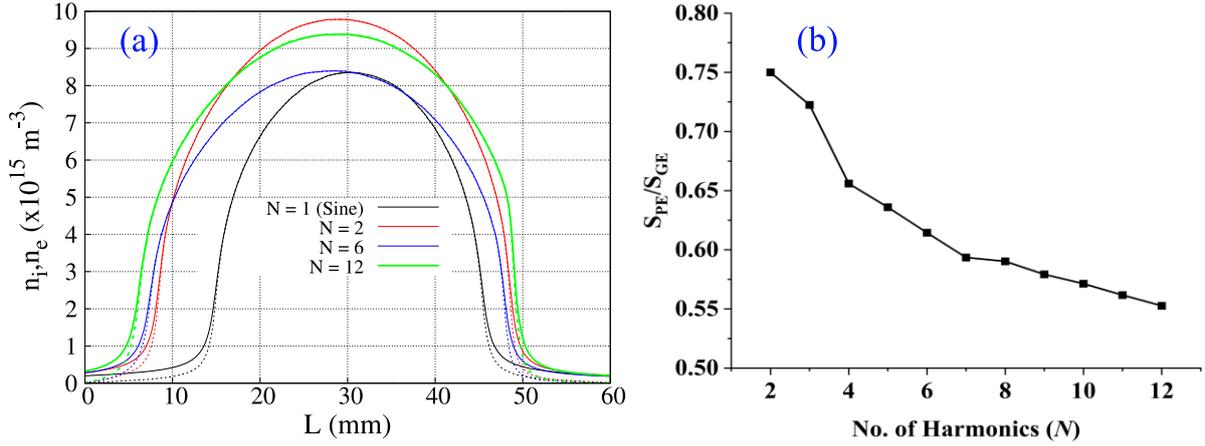

Fig. 2. (a) Time averaged spatial electron (dotted line) and ion density (continuous line) profile, and (b) ratio of sheath width at grounded electrode to powered electrode for different harmonics (*N*).

We further investigated the effect of *N* on the plasma density and ion flux towards powered and grounded electrodes. Figure 3 (a) shows the averaged peak plasma density in the discharge together with the ion flux at powered and grounded electrodes for different values of *N* contained in the sawtooth waveform. As shown, the plasma density is highest (~ $9.78 \times 10^{15}$ m$^{-3}$) for *N* = 2. For the sinusoidal case, the averaged peak plasma density is even lower ~ $8.36 \times 10^{15}$ m$^{-3}$ as the sheaths is bigger near the electrodes which decreases the bulk ionization and therefore the plasma density drops. As the number of harmonics increases, the average peak plasma density decreases, reaching a minimum of ~ $8.4 \times 10^{15}$ m$^{-3}$ for a critical value of *N* = 6, and then increases with a further rise in the number of harmonics. An investigation of the spatially averaged plasma density versus *N* shows similar variation i.e., it is first decreasing from ~ $5.58 \times 10^{15}$ m$^{-3}$ at *N* = 2 to ~ $4.82 \times 10^{15}$ m$^{-3}$ at *N* = 6 and then increasing with a further increase in *N* reaching to a maximum of ~ $5.78 \times 10^{15}$ m$^{-3}$ at *N* = 12. One of the crucial parameters in plasma processing is the ion flux towards the electrodes, which is closely related to plasma density. The simulation results predict that the ion flux variation with *N* at powered and grounded electrode (figure 3 (a)) is similar to that of plasma density. As shown in figure 3 (a), due to the larger sheath widths, the ion flux at the grounded electrode is consistently higher in comparison to the ion flux at the powered electrode. A similar trend of plasma density and ion flux is due to the fact that the ion flux ($\Gamma_i \propto nu$, where *n* and *u* are the plasma density and ion velocity (depends on the electron temperature) respectively at the sheath edge) is proportional to the plasma density. A similar behavior of both density and flux predicts that either electron temperature is not changing significantly with *N* or its effect on the ion flux is not significant for the present condition. Coming to the flux asymmetry, figure 3 (b) displays the ion flux ratio at the powered electrode to that of the grounded electrode. The simulation results show that the flux ratio is nearly constant, within 1%, for all the values of *N* from 2 to 12. These results suggest that the number of harmonics is affecting the ion flux arriving on the individual electrodes due to an asymmetry in the sheath width, however, the flux asymmetry is not changing with *N*. The results are in agreement with the previous observations in the voltage-driven *CCP* that predicts that the flux asymmetry vanishes when operating at a low gas pressure [53].

However, the ion flux variation with $N$ is nonlinear i.e., it first decreases, reaching a minimum value for a critical value of $N$, and then increases with a further rise in $N$. From a processing perspective, higher ion flux is essential for an enhanced processing rate thus both lower and higher values of $N$ are suitable for this purpose based on the flux requirements. Though, lower ion energies are also essential for damage-free processing. This will be discussed later in the manuscript.

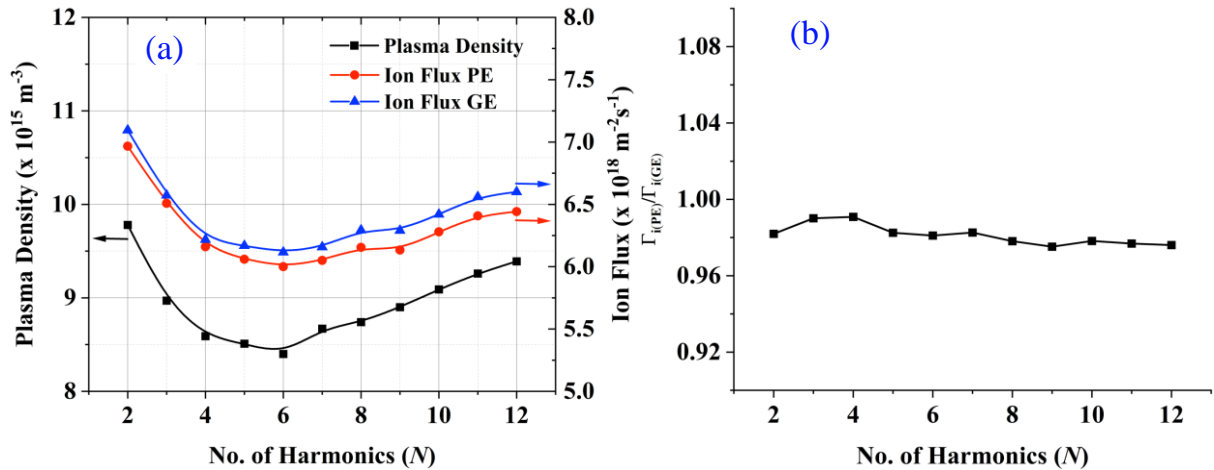

Fig. 3. (a) Averaged peak plasma density and ion flux at powered and grounded electrode, and (b) ratio of ion flux at powered electrode to grounded electrode for different harmonics (*i.e. N*).

The above results present that the number of harmonics contained in the saw-tooth waveform has a significant effect on the plasma density in the discharge and hence the ion flux varies accordingly on both powered and grounded electrodes. In order to understand the effect of $N$ on plasma density/ion flux, we have investigated the electron heating and ionization collision rate in the discharge. Figure 4 displays the total averaged electron heating and ionization collision rate in the discharge for different values of $N$. It is observed that the total averaged electron heating is following the same trend as the peak electron density presented in figure 3 (a). A minimum in the total electron heating is observed at $N=6$, which coincides with the electron density minima (figure 3 (a)). On both sides of the critical value of $N$, the total electron heating increases. The trend of ionization collision rate plotted in figure 4 presents the same variation as electron heating versus $N$. This suggests that electron heating is mostly being consumed in the production of electron-ion pair through the ionization process, and thus the plasma densities (figure 3 (a)) vary accordingly.

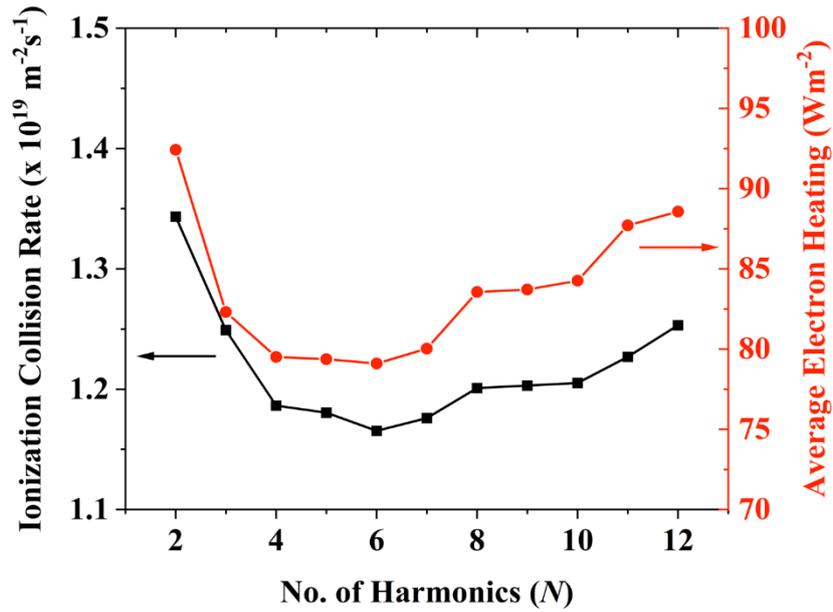

Fig. 4. Total ionization collision rate and electron heating in the discharge for different values of $N$.

For a detailed understanding, we further investigated the profile of electron heating in the discharge. Figure 5 shows the time-averaged electron heating ($<J.E>$) distribution in the discharge. In figure 5 (a), the profiles are plotted up to the critical values of $N = 6$, whereas above the critical $N$, the profiles are plotted in figure 5 (b). As displayed in figure 5, due to the low pressure consider the maximum heating is observed near the sheath edge and the bulk heating is nearly negligible. In the case of $N = 1$ *i.e.*, for the sinusoidal case, the time-averaged electron heating is symmetric, and its magnitude is ~100 W/m$^2$. However, the bulk plasma length in the sinusoidal case is smaller due to which ionization probability is low, and therefore plasma density is also lower in comparison to the sawtooth-like waveform. As $N$ increases, a clear asymmetry in electron heating is observed at grounded and powered electrodes. For up to the critical values of $N$ (≤6), the electron heating peak near electrodes is decreasing and the heating area inside the sheath region remains nearly constant. This confirms the total heating drops and hence the electron density decreases from $N = 2$ to $N = 6$. Furthermore, no significant electron cooling is observed near the sheath edge. Above $N = 6$, the electron heating peak is decreasing slightly or remains constant, however, the heating region (total area) inside the sheath near to the grounded electrode is increasing (i.e., between 50 mm to 60 mm) and thus the overall heating is increasing. Another important effect to be noticed here is the appearance of negative electron heating (figure 5 (b) zoomed picture) *i.e.*, electron cooling in the vicinity of the sheath edge (between 47 mm to 49 mm) at grounded electrode for different values of $N$ above 6. This electron cooling near the sheath edge drives the electron beam into the bulk plasma, which has been observed previously in *VHF* capacitive discharges [9-10,13-14,70]. An increase in the negative electron heating region suggests that the electron beams are getting more energetic for the higher values of $N$. Due to the low-pressure condition, these electron beams penetrate deeply into the bulk plasma and create high plasma density due to the ionization through highly energetic electron beams confinement from the opposite sheath edge [9-10,13-14,70].

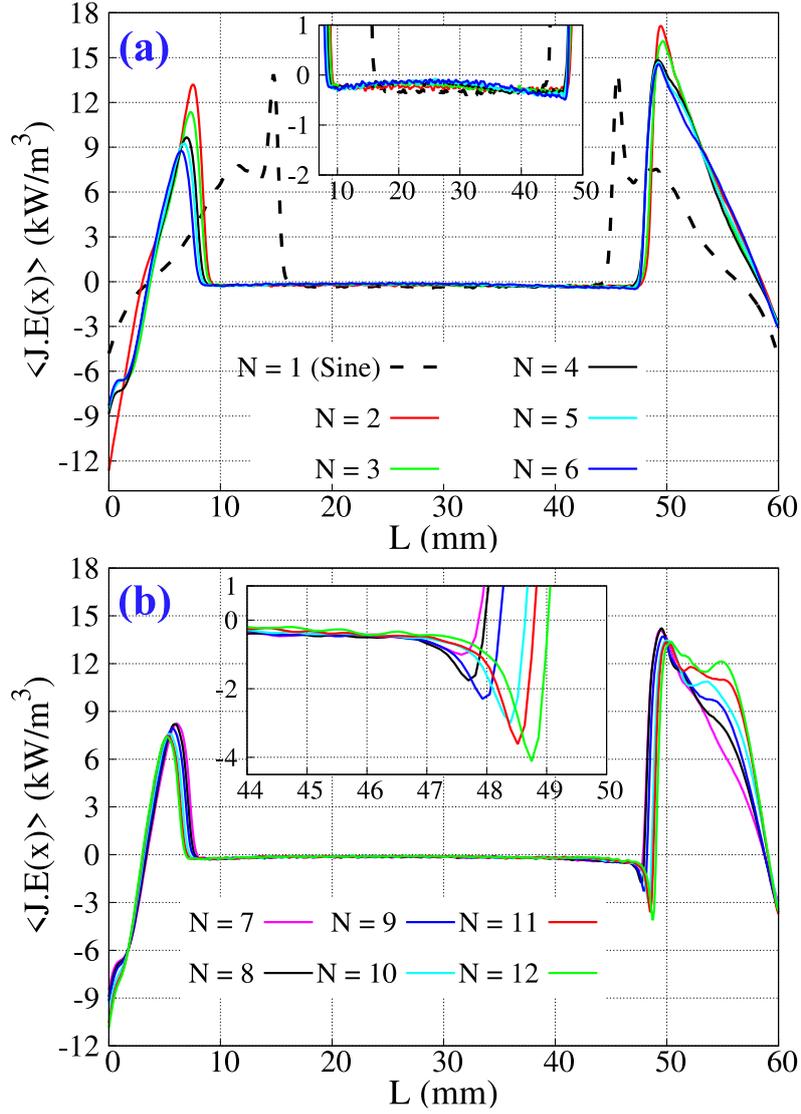

Fig. 5. Spatial profile of time averaged electron heating in the discharge for (a) $N = 1$-6 harmonics and (b) $N = 7$-12 harmonics.

Figure 6 shows the spatio-temporal evolution of the ionizing collision rate for the sinusoidal case and 3 different values of $N$ i.e., $N = 2$, $N = 6$ (critical value), and $N=12$ (highest value). The ionization rates were averaged over the last 600 *RF* cycles after reaching the steady state solutions. As displayed, the ionizing collisions are nearly symmetric for the sinusoidal case (figure 6 (a)), and the maximum occurs near to the expanding sheath edge at both electrodes. A single broad ionization beam is generated near the expanding sheath edge, penetrating the bulk plasma due to low-pressure conditions and extending up to the opposite sheath edge. Due to symmetry, a similar ionization beam is created from the opposite sheath edge during the other half *RF* cycle. In both cycles, the ionization beam is terminating at the collapsing phase of the sheath edge onto the counter electrode. Thus, for the sinusoidal case, due to poor confinement of the energetic electrons, the overall plasma density is lower, sheaths are bigger on both electrodes and the bulk plasma length is smaller. For $N = 2$ (figure 6 (b)), the ionizing collision rate becomes asymmetric with a maximum near the grounded electrode when compared to the powered electrode. This effect is attributed to faster sheath expansion near the grounded electrode because of the temporal asymmetric waveform and thus producing a higher ionization rate. As one can see that the powered electrode sheath is also producing a weaker ionization collision rate. This is due to the confinement of energetic electrons produced from the grounded electrode sheath that is getting confined during interaction with the powered electrode sheath. Thus, the overall plasma density is higher in the discharge at $N = 2$. For $N = 6$ (figure 6 (c)), the ionizing collision is mostly appearing near the grounded electrode due to the larger sheath width with almost negligible ionization collision rate near the powered electrode. Additionally, the peak ionization collision rate is lower in comparison to $N = 2$. Therefore, the plasma density is lowest for $N = 6$. At $N = 12$ (figure 6 (d)), a similar effect is seen as in the case of $N = 6$, however, now multiple ionization beams are observed from near to the expanding sheath edge at the grounded electrode. This multiple ionization beams generation is due to the multiple electron interaction with expanding sheath edge that triggers multiple beams of electrons. During these interactions, the instantaneous sheath edge position is modulated with high-frequency oscillations. Similar effects have been observed in capacitive discharges operating at a very high frequency [9-10,13-14,70]. These multiple ionization beams support higher plasma density in the discharge that is observed at $N = 12$.

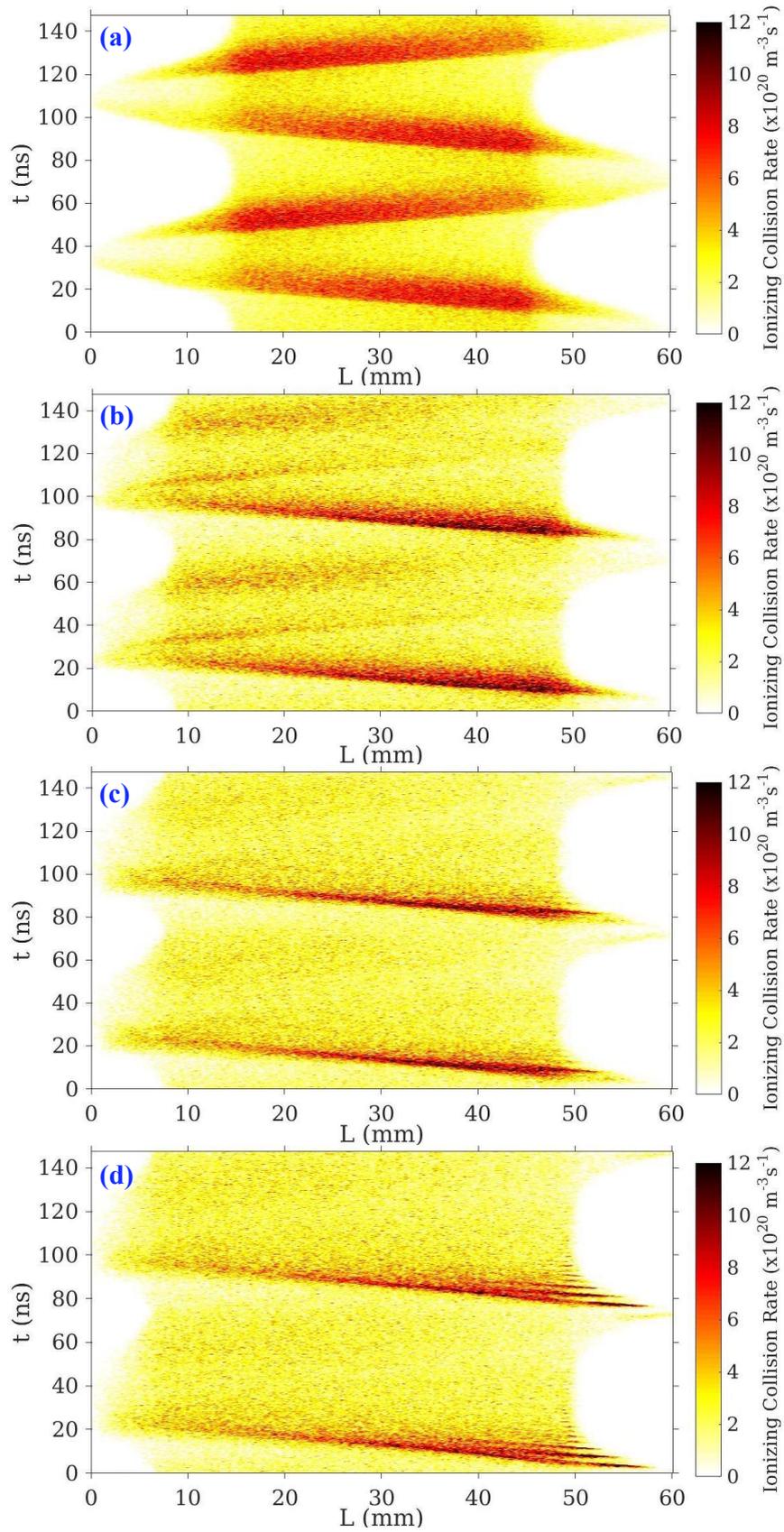

Fig. 6. Spatio-temporal ionizing collision rate in the discharge for (a) sinusoidal, (b) $N = 2$, (c) $N = 6$ and (d) $N = 12$.

The present trend in the ionizing collision rate and ion flux versus *N* is in contradiction with the voltage-driven simulation at higher pressure [54]. One of the reasons for this discrepancy is the operating pressure. In the present simulation, the operating gas pressure is low, and the electron beams produced from the expanding sheath edges are penetrating into the bulk plasma. Therefore, the ionization processes are not localized near the sheath region albeit extending into the bulk plasma, which is responsible for a symmetric ion flux. The ionizing collision rate is proportional to the electron heating and therefore the flux follows a similar trend versus *N*. An increase in the plasma density after a critical value of *N* (i.e., *N* = 6) is due to the multiple electron beam generation from near to the expanding sheath edge at the grounded electrode and the presence of higher harmonics on the instantaneous sheath edge position. Asymmetry in the sheath width at the powered and the grounded electrode is observed which might affect the ion energies arriving at the electrode surfaces, which is discussed in the following section.

One of the most important parameters in the plasma processes is the Ion Energy Distribution Function (*IEDF*) at the electrode surfaces. We investigated the *IEDF* at both the powered and grounded electrodes. Figure 7 shows the normalized *IEDF* at powered ((a), (b), and (c)) and grounded ((d), (e), and (f)) electrodes for sinusoidal waveform and for different values of *N*. In figure 7, different energy ranges are plotted on the *x- axis* for the proper visualization of *IEDF*. In figures 7 (b) and 7 (e), *IEDFs* are presented up to the critical value of *N* i.e., up to *N*=6, and figures 7 (c) and 7 (f) show *IEDFs* for *N* > 6. The values of peak energies are marked in the figures. The results show bi-modal *IEDFs* with a 75 eV energy spread (*ΔE*) for sinusoidal waveform and below 50 eV for *N*>2. The bi-modal shape of *IEDF* is attributed to the low-pressure conditions due to which collisionless ion transit occurs through the *RF* sheath. Furthermore, the energy spread of the bi-modal peaks depends on the ratio of ion transit time ($\tau_i$) through *RF* sheaths with period ($\tau_{RF}$) [75]. For the sinusoidal case, the calculated ratio of ion transit time to *RF* period is ~10 using the analytical expression defined in ref [75]. In the above calculation, average the sheath width is ~14 mm, and the mean sheath voltage is ~600 V. Thus, the ion will take many *RF* cycles to transit the sheath and responds to the average sheath voltage. This gives an energy spread of ~80 eV for a high-frequency regime ($\tau_i/\tau_{RF}$>>1), which is in agreement with the predicted simulation results (75 eV, figure 7 (a) and (d)). Furthermore, due to the larger sheath voltage and sheath width, the energy separation between the bi-modal peaks is higher at the grounded electrode in comparison to the energy separation between the bi-modal peaks at the powered electrode. The ratio of energy separation at the grounded to the powered electrode is ~1.53 at *N* =2, which increases to 2.05 at *N* =12 due to an increase in the discharge asymmetry with *N*.

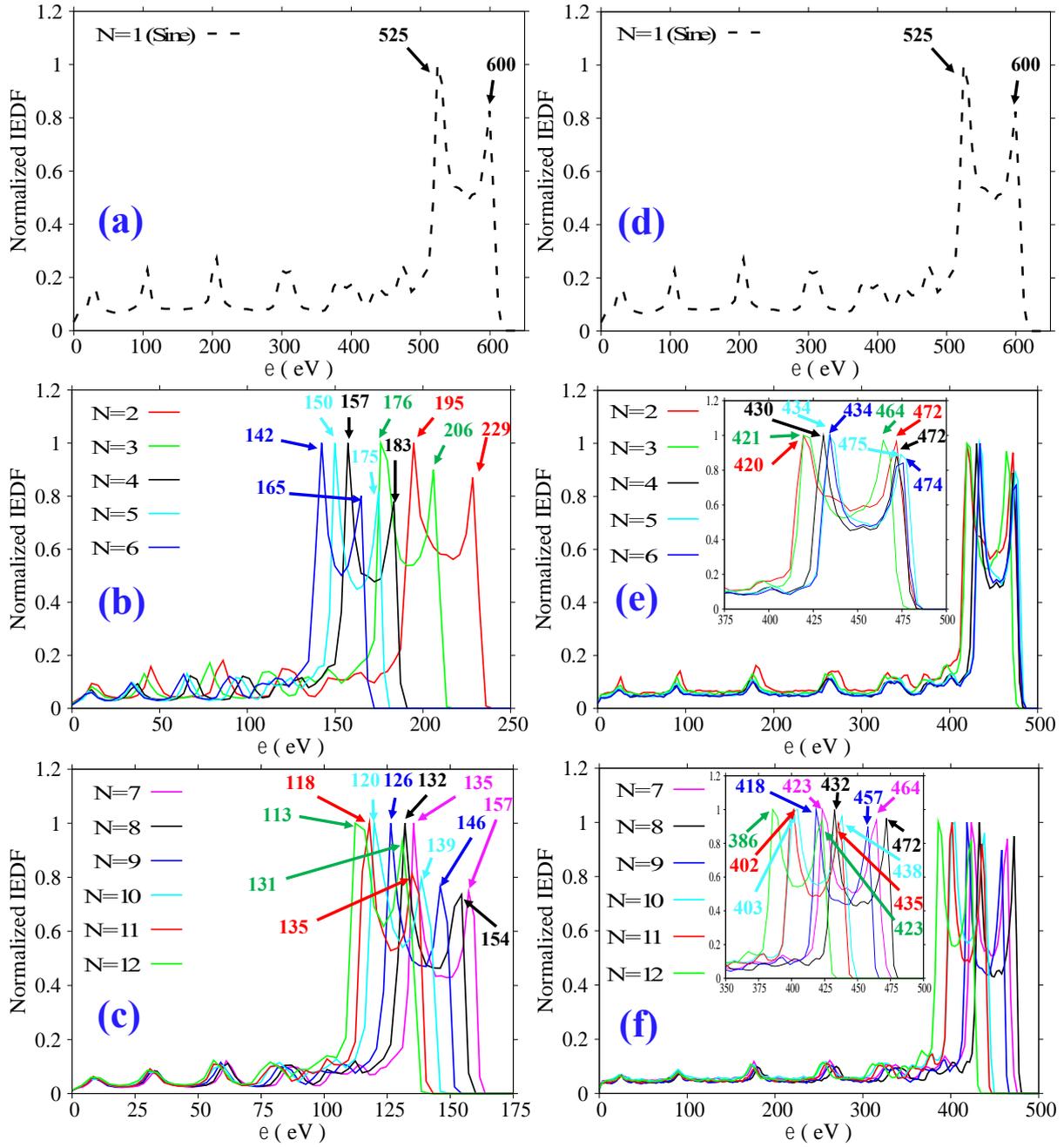

Fig. 7. Normalized Ion Energy Distribution Function (*IEDF*) at powered ((a), (b) and (c)) and grounded ((d), (e) and (f)) electrodes for sinusoidal waveform ((a) and (d)) and for different values of *N* (i.e., 2-6 in (b) and (e); 7-12 in (c) and (f)).

Regarding energy asymmetry, the results show higher ion energy at the grounded electrode when compared to the powered electrode. Figure 8 shows the systematic variation in the mean and maximum ion energy at the power electrode, and the ratio of average ion energies at the powered to the grounded electrode for different values of *N*. As displayed in figure 8, both mean and maximum ion energy at the powered electrode is decreasing with an increase in the value of *N*. This is due to the fact that the voltage drop across the sheath is decreasing with an increase in the number of harmonics. As displayed in figure 8 (right-hand Y axis in red), the ion energy asymmetry ($<\varepsilon_i>_{PE}/<\varepsilon_i>_{GE}$) increases with *N*. For the current values of *N* from 2-12, a ~40 % increase in asymmetry is observed. This is consistent with the sheath ratios plotted in figure 2 (b) where an asymmetry was observed in the sheath width. Thus, for the low-pressure case, it is concluded that the asymmetry due to a change in the number of harmonics of a sawtooth-like waveform is appeared only in the ion energy, whereas the ion flux asymmetry (figure 3 (b)) remains unaffected.

The *IEDF* at the electrode surface is determined by the potential difference between the plasma and the electrode. An investigation of the *DC* self-bias along with the spread in the bi-modal energy peaks (*ΔE*) is presented in figure 9. As displayed in figure 9, the plasma potential with respect to the *DC* self-bias is decreasing with an increase in the value of *N*. These values correspond to the mean ion energies arriving at the powered electrode shown in figure 8. A decreasing trend of ($V_p - V_{DC}$) is attributed to the formation of *DC* self-bias onto the powered electrode, which is changing at a faster rate in comparison to the plasma potential and therefore the potential difference between the plasma and the electrode decreases with *N*. The decrease in the ion energy is of primary importance in many plasma processes and therefore higher values of *N* are suitable for this purpose. Along with the lower ion energy, higher values of *N* is also suitable for a less spread ion energy distribution. As shown in figure 9, the energy spread in the bi-modal energy peaks (*ΔE*) is decreasing with an increase in the value of *N*. For the current chosen range (*N* = 2 – 12), the value of *ΔE* is decreased from 34 eV to 17 eV *i.e.*, a 50% drop in the energy spread is observed. A decrease in the value of *ΔE* is on the one hand due to an increase of high-frequency content in the driving waveform that drives *IEDF* from a bi-modal shape to a narrow distribution as observed in the very high frequency *CCP* discharges [10]. The high-frequency component on the instantaneous sheath edge position is clearly observed as shown in figure 6 (d). On the other hand, the energy separation (*ΔE*) is proportional to the sheath voltage, which is also found to decrease with *N,* and therefore *ΔE* decreases.

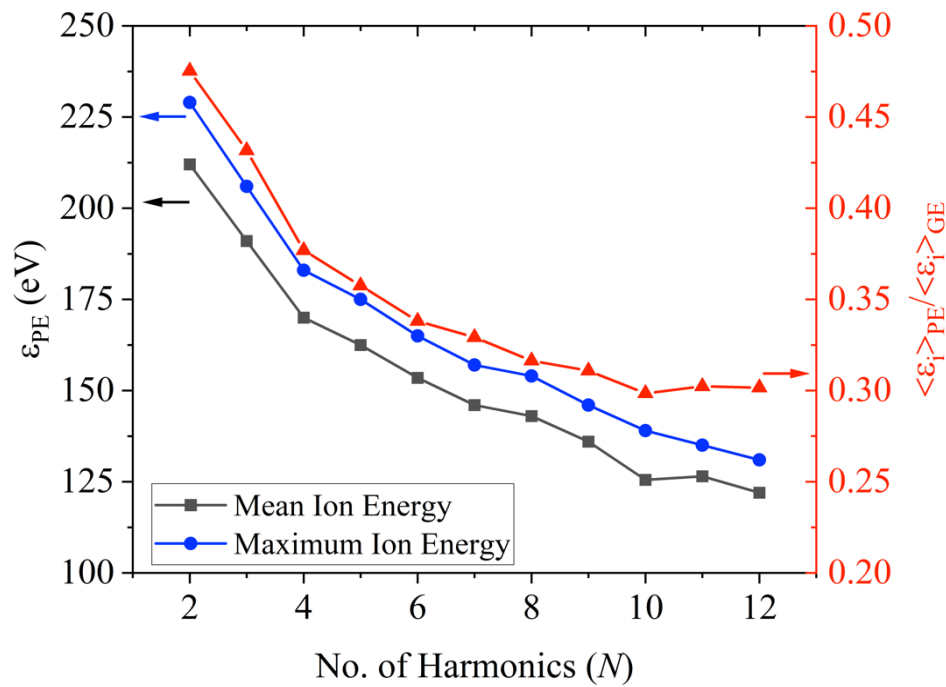

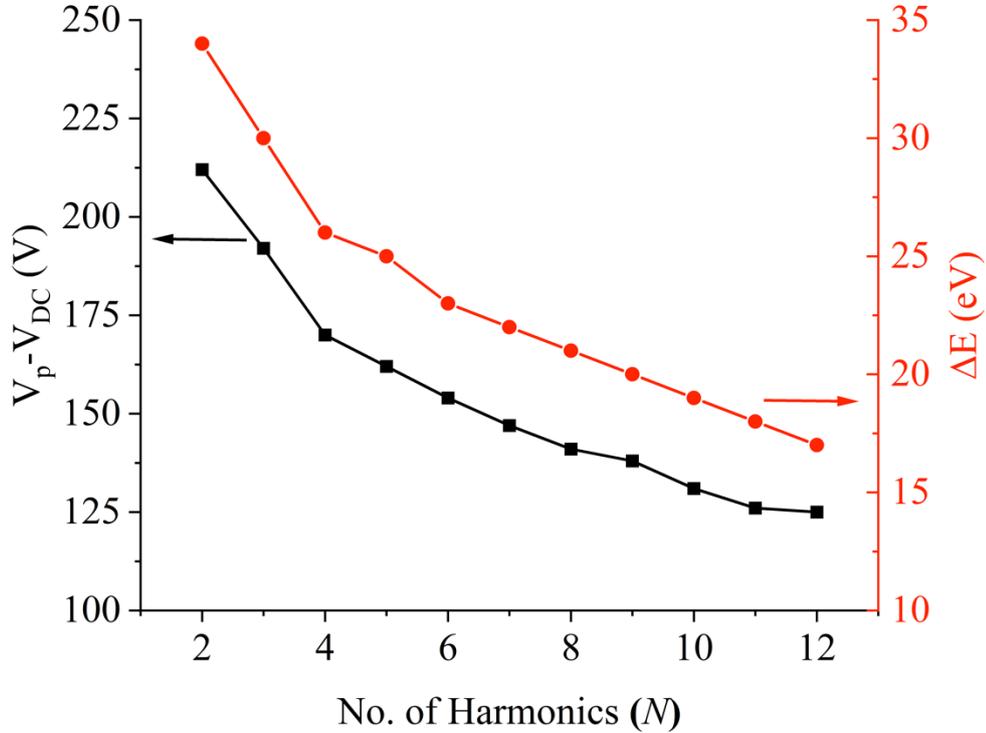

Fig. 8. Mean (black line and squares) and maximum ion energy (blue line and circles) at the powered electrode, and ratio of mean ion energy at powered and grounded electrode (red line and triangles) for different harmonics (*N*).

Fig. 9. Plasma potential with respect to *DC* self-bias at powered electrode and energy separation (*ΔE*) between bi-modal peaks for different harmonics (*N* = 2-12).

## IV. Summary and Conclusions

In a low-pressure *CCP* discharge, we have studied the effect of number of harmonics contained in the sawtooth-like waveform on the plasma asymmetry, electron-sheath heating, and *IEDFs* on the electrode surfaces. By using a *1D-3V PIC* simulation technique, we have observed that the plasma density profile becomes asymmetric with an increase in *N*. The sheath near the powered electrode decreases at a much faster rate than in comparison to the sheath near the grounded electrode and therefore the asymmetry in the sheath ratio at powered and grounded electrodes increases with *N*. A nonlinear trend in the peak plasma density is observed *i.e.*, it first decreases, reaching a minimum value for a critical value of *N,* and then increases with a further rise in *N*. A similar trend is observed in the ion flux at the powered and grounded electrode, ionizing collision rates, and time-averaged electron heating rates in the discharge. A study of the spatio-temporal evolution of the ionizing collision rates predicts that the ionization asymmetry increases with *N*. Due to the higher sheath width on the grounded electrode, it produces more ionization near it when compared to the powered electrode sheath. At higher values of *N*, we observe high-frequency modulation on the instantaneous sheath edge position on the grounded electrode due to which multiple ionization beams are generated. These multiple beams drive higher density in the discharge as the value of N increases above the critical value.

Due to low-pressure conditions, the time-averaged electron-sheath heating profile in the discharge shows maximum heating near the sheath edge position on the powered and grounded electrodes. For up to the critical values of $N$ ($\leq 6$), the peak electron heating near the sheath on both electrodes decreases, and no significant electron cooling is observed in the bulk plasma. For $N \geq 6$, the overall sheath heating area increases along with a strong electron cooling region near the grounded sheath edge position. This electron cooling region is the primary energy source for the electron and ionization beam trigger into the bulk plasma.

A systematic study of the *IEDFs* at the electrode surfaces shows a bi-modal shape for different values of $N$ up to the 12$^{th}$ harmonic. The energy separation (*ΔE*) of the bi-modal energy peaks decreases with a rise in the value of $N$ suggesting that the higher $N$ supports narrow *IEDFs*. Due to temporal asymmetry, *DC* self-bias is generated on the powered electrode, and the difference between plasma potential and *DC* self-bias decreases almost linearly with an increase in $N$. The corresponding mean and maximum ion energy arriving at the powered electrode also decreases. With increasing $N$, a clear asymmetry (~40%) in the mean ion energies at powered and grounded electrodes is observed. The simulation results conclude that, at low gas pressure, the higher value of $N$ is suitable for an enhanced ion energy asymmetry and narrower distribution, however, flux asymmetry is not generated.

**Acknowledgement**

This work is supported by the Science and Engineering Research Board (SERB) Core Research Grant No. CRG/2021/003536.